\documentclass[amsmath,amssymb,aps,superscriptaddress, twocolumn, print]{revtex4-2}

\usepackage[english]{babel}
\usepackage[utf8x]{inputenc}
\usepackage{amsmath}
\usepackage{graphicx}
\usepackage[colorinlistoftodos]{todonotes}
\usepackage{xcolor,soul}
\definecolor{cerclegreen}{HTML}{139F46}

\makeatletter
\renewcommand{\fnum@figure}{FIG. \thefigure}
\renewcommand{\fnum@table}{TABLE \thetable}
\makeatother

\begin{document}

\title{Thermoplasmonic Nanomagnetic Logic Gates}

\author{Pieter Gypens}
\affiliation{Dept. of Solid State Sciences, Ghent University, Belgium}

\author{Na\"emi Leo}
\affiliation{CIC nanoGUNE BRTA, E-20018 Donostia-San Sebastian, Spain}

\author{Matteo Menniti}
\affiliation{CIC nanoGUNE BRTA, E-20018 Donostia-San Sebastian, Spain}

\author{Paolo Vavassori}
\affiliation{CIC nanoGUNE BRTA, E-20018 Donostia-San Sebastian, Spain}
\affiliation{IKERBASQUE, Basque Foundation for Science, E-48009 Bilbao, Spain}

\author{Jonathan Leliaert}
\email{jonathan.leliaert@ugent.be}
\affiliation{Dept. of Solid State Sciences, Ghent University, Belgium}

\begin{abstract}
Nanomagnetic logic, in which the outcome of a computation is embedded into the energy hierarchy of magnetostatically coupled nanomagnets, offers an attractive pathway to implement in-memory computation. This computational paradigm avoids separate energy costs associated with transporting and storing the outcome of a computational operation. Thermally-driven nanomagnetic logic gates, which are driven solely by the ambient thermal energy, hold great promise for energy-efficient operation, but have the disadvantage of slow operating speeds due to the lack of spatial selectivity of currently-employed \textit{global} heating methods. As has been shown recently, this disadvantage can be removed by employing \textit{local} plasmon-assisted photo-heating. Here, we show by means of micromagnetic and finite-elements simulations how such local heating can be exploited to design reconfigurable nanomagnetic Boolean logic gates. The reconfigurability of operation is achieved either by modifying the initialising field protocol or optically, by changing the order in which horizontally and vertically polarised laser pulses are applied. Our results thus demonstrate that nanomagnetic logic offers itself as a fast (up to GHz), energy-efficient and reconfigurable platform for in-memory computation that can be controlled via optical means.
\end{abstract}

\maketitle

\section{Introduction}
The success of computing in the past decades has been based on the development of charge-based electronics with a von-Neumann architecture, which separates the domains of computation and data storage. However, due to the increasing demand in analysis of ``big data'' in recent years, alternative computation architectures and physical platforms are explored for their potential improvements in energy efficiency and data access~\cite{2018manipatruni, 2018DiVentra, 2020Sebastian}. Magnetic phenomena such as domain walls~\cite{ALL-05, BAR-06, FRA-12a, LUO-20}, skyrmions~\cite{ZHA-15}, and magnetic tunnel junctions~\cite{NEY-03, WAN-05, LEE-07a} have been proposed to lie at the basis of new computing paradigms to complement current technology which relies on complementary metal-oxide-semiconductors (CMOSs). Another candidate that holds great promise because of its non-volatility and its potential to operate with low dissipative power losses is nanomagnetic logic (NML)~\cite{COW-02, CSA-02a, CSA-02b, CSA-04, IMR-06, PUL-11, CAR-11, GU-15}. It makes use of single-domain nanosized islands with shape anisotropy such that the magnetisation has two preferential orientations, allowing to encode bits $0$ and $1$. This state information is communicated to other islands by magnetic stray fields, which can be tailored by placing the islands in a suitable geometrical arrangement in order to implement logical functionalities, e.g, the three-input majority gate~\cite{IMR-06} and NAND gate~\cite{GYP-18, GYP-21}.

The fundamental principle of nanomagnetic logic is that the logically correct output of the gate corresponds to a low-energy state of the interacting ensemble. This state needs to be reached by a series of successive reversals of individual nanomagnets, which are triggered by application of an external clocking field or via spontaneous thermal activation. While plenty of works considered the design of nanomagnetic logic circuits with the logically correct states corresponding to the minima in the energy landscape~\cite{IMR-06, ARA-18, GYP-18, ARA-19}, less emphasis has been put on their dynamical behaviour. From a technological point of view, however, the operation mode dictates both the speed and the energy efficiency of the gate.

Operating modes based on an applied clocking field always require an external power source. The efficiency of such schemes has increased tremendously thanks to the advent of new ways to generate the clocking field. Examples that emerged over the past decade are for instance the use of the spin-Hall effect~\cite{BHO-14} or strain-mediated switching~\cite{ATU-10, d-16} to toggle the state of the magnetic islands. Nevertheless, the energy efficiency of thermally-driven systems which rely solely on the ambient thermal energy can not be matched~\cite{CAR-11}.

The drawback of thermally-driven nanomagnetic logic circuits which relax at a constant temperature ~\cite{CAR-11, CAR-12, ARA-18, ARA-19} is their slow operating speed (typically several orders of magnitude lower than that of clocked systems, up to several hours). This slow operating speed is due to the fact that the relaxation towards the ground state occurs through a Brownian motion of high-energy defects which need to find their way out of the system. Thermal annealing, i.e. slowly cooling the \emph{entire} system, provides the necessary driving force to oppose such a random walk and to drive a calculation in the forward direction, i.e. to a state of lower energy~\cite{GYP-18}. However, this approach yields operating speeds that cannot compete with those obtained by clocking. In addition, a \emph{global} heating protocol limits the energy efficiency as it heats the entire device, impeding the integration with CMOS which could be an intermediate step for applications. Moreover, a global heating protocol is not effective for extended systems which consists of hundreds of islands, because it lacks spatial selectivity.

To address the problems of reliable and localised operation, it would be advantageous to control the temperature of individual nanomagnets. Such \emph{local} heating has been reported recently by integrating a thermoplasmonic heater with a ferromagnetic nanoisland~\cite{PAN-19}. Such thermoplasmonic heaters, which are typically used in biomedical applications such as innovative cancer therapy and temperature-activated drug delivery~\cite{o-14, JOR-16, LI-18}, efficiently convert incident laser light into heat when the laser resonantly excites the localised surface plasmons (LSPs), i.e., collective oscillations of the conduction electrons at a metal-dielectric interface driven by the electromagnetic field. Given the nanostructure and substrate materials, the wavelength at which the LSP is excited depends on the size of the heater. For elliptical heaters, the excitation furthermore occurs at different wavelengths depending on the light polarisation.

This polarisation dependence provides a tool to control the temperature of the hybrid ferromagnet-heater nanostructure and thus the magnetic properties and switching behaviour of the ferromagnetic component~\cite{PAN-19}, making it possible to engineer a relaxation pathway that guides a NML gate directly to the desired ground state. The idea of engineering relaxation pathways in square artificial spin ice (ASI) has previously been explored by Arava~\cite{ARA-18, ARA-19}, who designed an NML gate that relaxes from a field-set state to a low-energy state via a pre-known series of switching events. There, the relaxation pathway is promoted by global heating and based on a monotonous energy decrease.

In this work, we show a NML gate design in which \emph{local} thermoplasmonic heating is exploited to drive the system reliably and deterministically towards the lowest energy state on nanosecond time scales. We thereby effectively show the oft-mentioned potential of ASI in logical applications~\cite{JEN-18}. The reconfigurable gate behaviour (AND or OR) can be controlled either by applied fields upon initialisation, or even dynamically during operation by changing the laser illumination. By means of micromagnetic and finite-elements simulations, we demonstrate a high reliability of the gate output and discuss physical considerations to take into account for successful gate design. The typical energy cost for a computational operation based on our approach of light-controlled nanomagnetic logic is in the pJ range~\footnote{For the specific and illustrative case reported here, the optical energy cost is about 100 pJ.}, whereas it is in the aJ range for ultimate CMOS technology~\cite{MAN-18}. Therefore, our approach brings about the additional advantages of non-volatile information storage, optical synchronisation, and potential reconfigurability at an acceptable loss in efficiency.

\section{Plasmon-assisted photo-heating of nanomagnets}

In order to efficiently heat a nanomagnet using laser light, one can combine it with a thermoplasmonic heater~\cite{PAN-19}, which are commonly made of gold because of its sharp, intense plasmon resonance peaks and its chemical stability~\cite{2007Maier, 2017Baffou}. From a practical viewpoint, both components should have the same size, which for ferromagnetic islands in logic applications is typically a few tens of nanometers. Thermoplasmonic heaters of such a size have a plasmonic eigenfrequency corresponding to near-infrared to optical light when embedded in a low-refraction-index dielectric (e.g., an air/glass surface as considered here). Thus, illuminating a ferromagnet-heater nanostructure (e.g., permalloy-gold) with an optical laser of selected wavelength to resonantly excite the LSPs of the heater increases the temperature of the entire island, as the absorbed photon energy will be converted into heat. In this study, we consider a temperature increase amounting to hundreds of Kelvin, as modulated by the pulse duration and the beam fluence (i.e., the power per illuminated area [W/m$^2$]). The focal position of the beam makes it possible to heat a specific area, which is not only energetically efficient but also allows us to target individual gates in systems where multiple gates are integrated, without affecting neighbouring circuits.

In addition to the spatial selectivity, island-specific plasmonic heating can be achieved via the polarisation-dependent optical absorption cross section displayed by elliptical nanoheaters. Using the appropriate wavelength, ellipsoidal islands heat up substantially when the polarisation is along their geometrical long axis. In contrast, light of the same wavelength hardly induces any temperature rise when the polarisation is along the short axis~\cite{PAN-19}. As a result, tuning the polarisation enables some islands to efficiently absorb the incident light, whereas the others hardly show any heating.

Island-selective plasmonic heating offers a route to control the magnetisation reversals that can occur in arrays for which the geometrically long axis of the constituent islands are not aligned (e.g., an artificial square spin ice lattice). Such an array can be regarded as consisting of two sublattices. The islands of the {\it hot} sublattice have their long axis aligned with the beam polarisation, hence exhibiting strong thermoplasmonic efficiency, such that the energy barrier that must be overcome in order to switch lowers due to the the reduced saturation magnetisation. In contrast, the islands whose short axis is along the beam polarisation belong to the {\it cold} sublattice, with the magnetisation direction remaining fixed upon irradiating.

To prevent the cold sublattice from being heated via thermal diffusion (i.e., dissipation of the heat generated in the hot sublattice via the substrate), short pulses from pico- to nanosecond time scales should be used~\cite{PAN-19}. These pulses allow for the excitation of plasmonic resonances (in the gold component, on fs time scales, and internal temperature equilibration via electron-phonon scattering on ps time scales) and subsequent heating of the ferromagnetic component. Due to the shortness of the pulse, the heat generated in the nanostructure enables temperature increases up to several hundred Kelvins, while little heat is transferred to the substrate. As the thermal diffusion is limited to the very close surrounding of the nanoheater and does not reach the next neighbouring nanostructure, a collective and global heating of the entire system can be avoided~\cite{2013Baffou}.

To implement a reliable nanomagnetic logic gate operation, the beam fluence of these short pulses is tuned to a trade-off between {\it i)} a temperature rise sufficiently high to obtain a reasonable switching probability, and {\it ii)} a temperature rise sufficiently low to keep the magnetic interaction strength sufficiently strong.

\begin{figure*}
\centering{
    \includegraphics[width=0.75\textwidth]{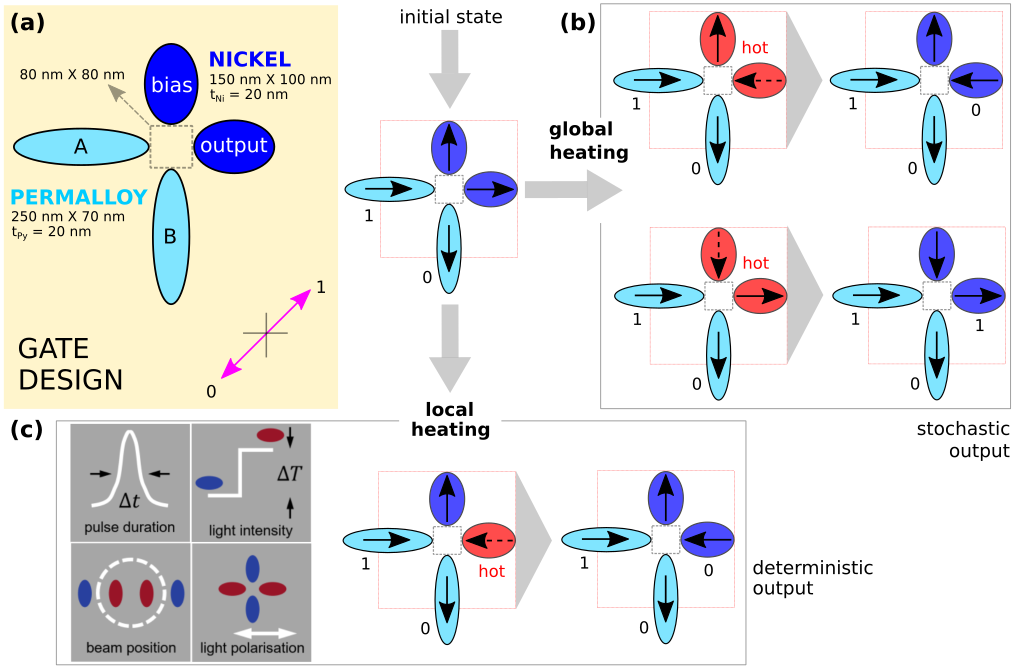}}
    \caption{(a) A thermoplasmonic NML gate consisting of four nanoislands arranged in a square vertex, similar to artificial square spin ice, where bit $0$ and $1$ correspond to a magnetisation pointing to left/down and right/up, respectively, as indicated by the pink arrows. The input islands, labeled A and B, are configured to have a fixed magnetisation direction. The ferromagnetic islands are placed on top of gold nanoheaters with a thickness $t_{{\rm Au}} = 30$~nm. The substrate is made of glass. (b) In the case of a global heating scheme, each of the nickel islands -- which are more likely to switch due to their smaller saturation magnetisation -- can relax, resulting in two possible final states and thus a stochastic output. (c) Deterministic gate operation is obtained when only the output island is heated, which can be achieved by polarisation-dependent heating of hybrid nanomagnet-plasmonic islands. Thermoplasmonic heating offers additional advantages for remote control, using the pulse length, intensity, and focal position as optical degrees of freedom, as shown in the inset.}
    \label{fig_gateDesign_global-local} 
\end{figure*}

\section{Gate design}

The operating principle of our thermoplasmonic NML gates is based on a relaxation from a high-energy state to the logically correct low-energy state through a series of magnetisation reversals, with the relaxation path being engineered by taking advantage of the island-selectivity of thermoplasmonic heating. Such a gate can be realised with four nanoislands arranged as the vertex of a square artificial spin ice, as shown in Fig.~\ref{fig_gateDesign_global-local}(a). Bit $0$ corresponds to a magnetisation pointing to the left or down, bit $1$ to a magnetisation pointing to the right or up, as indicated by the pink arrows in Figure~\ref{fig_gateDesign_global-local}(a).

The permalloy (Py, a nickel-iron alloy) input islands A and B have lateral dimensions of 250$\times$70~nm$^2$, while the vertical bias island and the horizontal output island are made of nickel (Ni) with lateral dimensions of only of 150$\times$100~nm$^2$. The thickness of all ferromagnetic islands is equal to $t_{{\rm FM}}=20$~nm and the edge-to-edge inter-island distance is 80~nm. The ferromagnetic islands incorporate thermoplasmonic heaters made of gold (Au) with a thickness of $t_{{\rm Au}} = 30$~nm, which are placed on a glass substrate (see Ref.~\cite{PAN-19} for a schematic representation).

It is worth emphasising that our design only behaves as a deterministic NML gate because of the island-specific heating. As illustrated in Fig.~\ref{fig_gateDesign_global-local}(b), global heating gives rise to two distinct relaxation paths, in which either the horizontal output island (top row) or the vertical bias island (bottom row) switches, yielding a stochastic output, as the magnetisation of the output can take the direction that corresponds to bit $0$ and $1$. In contrast, plasmon-assisted photo-heating makes it possible to select which island gets heated, such that only one relaxation path is energetically allowed and the output of the gate becomes deterministic [see Fig.~\ref{fig_gateDesign_global-local}(c)].

In addition to the practical realisation using state-of-art nanolithography, the materials and island sizes are chosen with a threefold objective in mind: to allow for {\it i)} a field protocol that can set the initial high-energy states, {\it ii)} optical selectivity to reverse the magnetisation of the bias or output island by heating it locally, without substantially increasing the temperature of the other islands (thereby keeping their magnetisation direction unchanged), and {\it iii)} magnetostatic interactions that are sufficiently strong to induce the desired logical behaviour.

\subsection{Initialisation: field protocol}
\begin{figure*}
\centering{
    \includegraphics[width=0.7\textwidth]{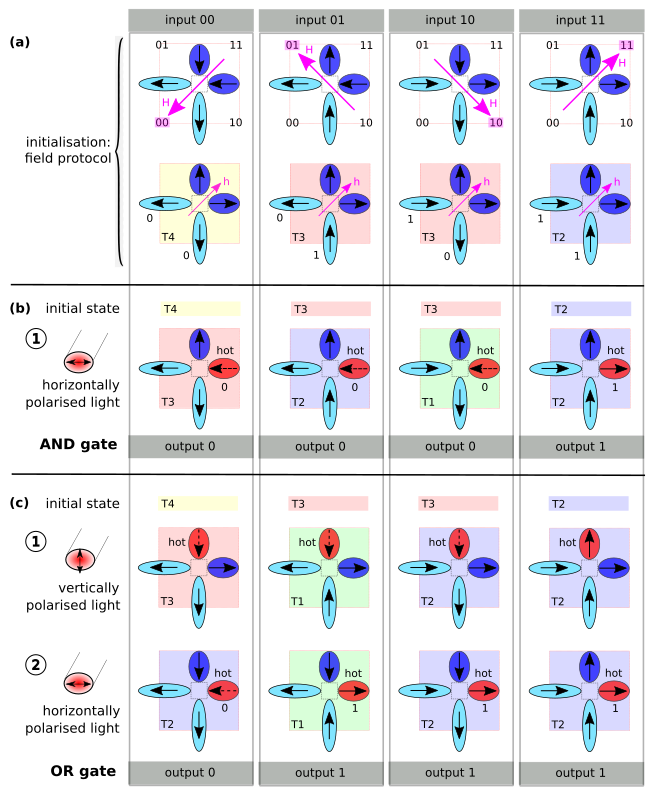}}
    \caption{Graphical representation of the operating principle of our thermoplasmonic gate, where the states are labeled according to the nomenclature of the energy hierarchy of artificial spin ice vertices with $E(T4) > E(T3) > E(T2) > E(T1)$. (a) The first step of the field protocol, shown in the top row, sets the input bits via the saturation with field $\vec{H}$. In the second step of the field protocol, a smaller field $\vec{h}$ is applied to set the initial high-energy states from which the gate will relax. (b) AND logic is induced by irradiating the entire gate with a laser beam of horizontally polarised light. Due to the preferential thermoplasmonic heating of the horizontal nickel island (output), the magnetisation of this island can reverse to a state of lower energy, as indicated by the dotted arrows. (c) OR logic is induced by irradiating the entire gate with a laser beam of vertically polarised light (which allows the switching of the vertical nickel island) and thereafter with horizontally polarised light (which allows the switching of the output island).}
    \label{fig_fieldprotocol} 
\end{figure*}

The field protocol, visualised in Fig.~\ref{fig_fieldprotocol}(a), is used to set the initial high-energy states, with the magnetisation of the input islands pointing in and out to encode bit $1$ and bit $0$, respectively, and with the magnetisation of the bias and output island always pointing outward from the vertex center. The input islands are initialised by applying a sufficiently large magnetic field $\vec{H}$ along the $00$, $01$, $10$, or $11$-direction, depending on the input configuration to be set. The field-saturated state thus obtained is subsequently subjected to a second, smaller, external magnetic field $\vec{h}$ which reverses the magnetisation of the nickel bias and output island to point outwards, but whose amplitude is too small to reverse the magnetisation direction of the permalloy input islands. The latter have a larger coercive field, due to the higher aspect ratio (3.6 versus 1.5) and zero-Kelvin saturation magnetisation (965~kA/m versus 615~kA/m~\cite{COE-10}).

\subsection{Gate operation: optical selectivity and magnetostatic interactions}
The configurations as initialised with the field protocol are high-energy states, labeled according to the conventional nomenclature of artificial spin ice vertices with energy $E(T4) > E(T3) > E(T2) > E(T1)$~\cite{WAN-06} (see Figs.~\ref{fig_sts}(a-b) for all state configurations and their energy). As shown in Figs.~\ref{fig_fieldprotocol}(b-c), the relaxation toward the low-energy state is governed by irradiating the entire gate with a laser beam. Horizontally polarised light heats the horizontal output island (allowing it to switch when energetically favourable) whereas vertically polarised light is applied to heat the vertical bias island. The input islands are configured to remain cold upon illumination such that their magnetisation remains fixed.

For input configurations $00$, $01$ and $10$, the island-specific heating is needed to control the state to which the system relaxes. For input configuration $11$ (fourth column of Fig.~\ref{fig_fieldprotocol}), no switching occurs because the reversal of the output or bias island would lead to a high-energy T3 state. The initial T2 state can relax to the T1 ground state without going through a T3 state only in the case of a double contemporary switching event. This might be possible via global heating (e.g., with light polarisation at 45~degrees or with circularly polarised light), although such a double contemporary switching event can be considered very unlikely.

To achieve the desired island-selective heating for input configurations $00$, $01$ and $10$, the difference in the polarisation-dependent LSP absorption cross sections should be as large as possible for the bias and output islands. Therefore, the long ellipse diameter is set at $d_{{\rm max}}=150$~nm such that the maximum of the plasmonic resonance occurs at a laser wavelength of $\lambda = 728$~nm for light polarisation along this axis. The length of the short axis is mainly determined by the condition that light polarisation parallel to this axis should result in a significantly lower LSP energy-absorption efficiency, putting an upper limit on $d_{{\rm min}}$. Using $d_{{\rm min}}=100$~nm, which allows reliable fabrication using nanolithography methods, the absorption cross sections are $4.1 \cdot 10^{-14}$m$^2$ and $0.7 \cdot 10^{-14}$m$^2$ for polarisation along the long and short axis, respectively (see Appendix~A for more details).

While the horizontal \emph{output} island gets heated upon illumination, the horizontal \emph{input} island A should remain sufficiently cold to keep its magnetisation direction fixed. A similar reasoning applies to the vertical \emph{bias} island and the vertical \emph{input} island B. Therefore, the axes of these input islands are chosen such that the LSP frequency does not match the frequency of the laser beam, i.e., that they exhibit poor thermoplasmonic efficiency. As detailed in Appendix~A, the absorption cross section of the 250$\times$70~nm$^2$ input island A is equal to $2.3 \cdot 10^{-14}$m$^2$ in the case of irradiation along the long axis, which is almost half compared to the output and bias island. The other input island B, having its short axis along the beam polarisation, is not heated at all, since the absorption cross section for irradiation along the short axis is 10 times smaller.

In addition, due to the higher Curie temperature of permalloy as compared to nickel (843~K versus 628~K), the saturation magnetisation of the input islands and hence their switching energy barrier are marginally affected by plasmonic heating; at the maximal light-induced temperature of about 600~K, the decrease of $M_{{\rm s}}$ is about only 25\% with respect to room temperature [see the temperature profiles of islands A and B shown in Fig.~\ref{fig_tempProfile}(d)].

The transitions shown in Fig.~\ref{fig_fieldprotocol}, on which the operation of our gate relies, involve the reversal of an island's magnetisation. Such a reversal occurs when the local temperature of the island approaches its Curie temperature $T_C$. For a reliable switching event from one state to another, the magnetostatic energy of both states must remain substantially different close to $T_C$ such that the transition probabilities are highly asymmetric, in the sense that only the transition from the high-energy to the low-energy state can take place.

Therefore, our gate is designed to have sufficiently strong magnetostatic interactions, by setting the thickness at $t_{{\rm FM}}=20$~nm to enlarge the magnetic volume, by placing the islands close with an inter-island distance of $R=80$~nm, and by using input islands made of permalloy because of its high saturation magnetisation. Since the bias and output island are heated up to $T_C$, their saturation magnetisation almost vanishes, making it less relevant for yielding strong interactions. 

\subsection{Reconfigurability}
\label{Reconfig}
Employing the field protocol of Fig.~\ref{fig_fieldprotocol}(a) and a horizontally polarised light pulse, our gate displays the behaviour of a logical AND gate [Fig.~\ref{fig_fieldprotocol}(b)]. However, the same gate design can be used as a logical OR gate by altering the second step of the field protocol, with the small external field $\vec{h}$ pointing to the bottom right instead of the top right, such that the magnetisation direction of the vertical bias island is initialised downwards instead of upwards, yielding a $T3$, $T1$, $T2$, and $T3$ as the initial state for input configuration $00$, $01$, $10$, and $11$, respectively. These states are then subject to horizontally polarised light, allowing the magnetisation of the output island to reverse from right ($\rightarrow$) to left ($\leftarrow$) when energetically favourable. The transition occurs for input configuration $00$ ($T3 \to T2$), but not for input configuration $01$ ($T1 \to T3$), input $10$ ($T2 \to T3$), and input configuration $11$ ($T3 \to T4$). As a result, only input configuration $00$ gives rise to output $0$, in accordance to a logical OR gate.

As shown in Fig.~\ref{fig_fieldprotocol}, reconfigurability can also be achieved in an optical manner, which is particularly advantageous to dynamically reprogram the logical operation even after its initialisation. The light protocol determines whether the gate displays the behaviour of a logical AND gate [Fig.~\ref{fig_fieldprotocol}(b)] or a logical OR gate [Fig.~\ref{fig_fieldprotocol}(c)]. The magnetisation of the output island settles into a direction which is consistent with AND logic by applying a horizontally polarised light pulse. In contrast, OR logic is observed in the case of a vertically polarised pulse followed by a horizontally polarised pulse. Thus, while it suffices to apply a single pulse to induce AND logic, the OR gate requires the subsequent application of both light polarisations. Considering the energy of operation and speed for an individual gate, it is therefore beneficial to use the field reconfigurability to induce OR logic, in which case only one pulse is needed.

The benefits of optical reconfigurability reach their full potential in more complex gates consisting of more nanoislands, where the multitude of possible transitions can be controlled by the light protocol. This would allow to reconfigure the gate in a way that cannot be achieved with fields.

\section{Gate performance}

\subsection{Simulations}
\begin{figure}
\centering{
    \includegraphics[width=0.45\textwidth]{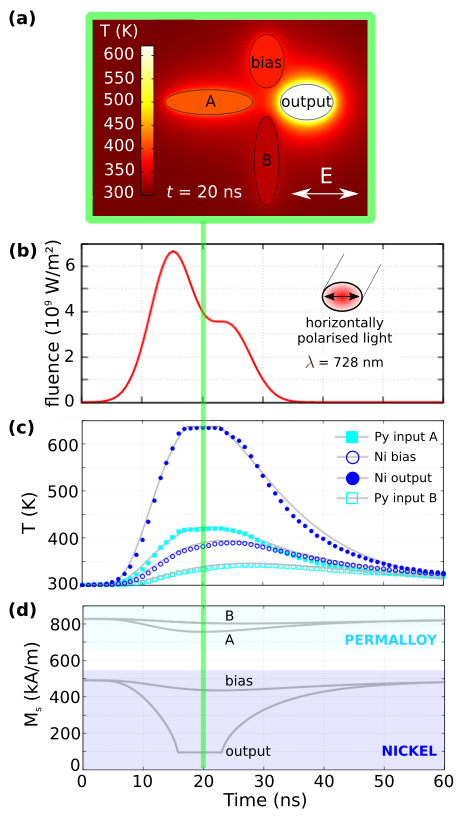}}
    \caption{(a) The temperature of the gate $t=20$~ns after the start of the laser pulse with light polarisation $\vec{E}$ along the horizontal direction (white arrow). (b) The beam fluence as a function of time when applying two Gaussian laser pulses of 7.8~mW focused on a circular 2D Gaussian spatial spot of 1 micron full-width at half maximum (FWHM). The horizontally polarised light pulses, with wavelength $\lambda=728$~nm, are 8.5~ns FWHM and separated by 9.5~ns. The amplitude of the second pulse is only 50\% compared to the first pulse. (c) The temperature of each island as a function of time due to the applied laser pulses. The temperature profiles (open and filled symbols) are simulated using COMSOL Multiphysics\textregistered~\cite{COMSOL}. A log-normal distribution has been fitted to the data (solid lines). (d) The saturation magnetisation of each island as a function of time.}
    \label{fig_tempProfile} 
\end{figure}

We use COMSOL Multiphysics\textregistered~\cite{COMSOL} to model the light-induced temperature increase in the hybrid ferromagnet-heater nanostructure [Fig.~\ref{fig_tempProfile}(a)]. Both the horizontally and vertically polarised light consists of two Gaussian light pulses, each one of 8.5~ns full-width at half maximum (FWHM) and separated by 9.5~ns, as shown in Fig.~\ref{fig_tempProfile}(b). The amplitude of the second pulse is only 50\% compared to the first pulse, where the pulse energy of the latter is equal to 70~pJ (yielding a net pulse energy of 105~pJ). This double-pulse excitation at a wavelength of $\lambda=728$~nm is chosen to obtain a temperature profile in which only the output island is heated close to its Curie temperature $T_C$ and remains at this temperature for a time of about 10~ns, as shown in Figure~\ref{fig_tempProfile}(c). This enables sufficient time for the nickel output island to reliably switch. A log-normal distribution, cut-off at the Curie temperature of the nickel output island ($T=625$~K), is fitted to the temperature profile $T(t)$ of each island.

These fitted temperature profiles $T_i(t)$ are used in the micromagnetic simulation program MuMax3~\cite{VAN-14a,LEL-17} to describe the time evolution of the respective saturation magnetisation $M_{{\rm s},i}(T(t))$, with $i$ denoting the A, B, bias or output island [see Fig.~\ref{fig_tempProfile}(d)]. The saturation magnetisation is equal to
\begin{equation}
    M_{{\rm s},i}(T(t)) = M_0 \big(1 - T_i(t) / T_C \big)^{\beta},
    \label{eq_Mstemp}
\end{equation} 
with $\beta = 0.35$~\cite{POR-13,PAN-19} and the Curie temperature being $T_C = 843$~K and $T_C = 628$~K for the permalloy and nickel islands, respectively. The theoretical value is used for the saturation magnetisation at zero Kelvin, i.e., $M_0 = 965$~kA/m (permalloy) and $M_0 = 615$~kA/m (nickel)~\cite{COE-10}. Plots of the saturation magnetisation of nickel and permalloy as a function of temperature can be found in Fig.~\ref{fig_msat-temp} (Appendix~B).

Our gate design, shown in Fig.~\ref{fig_gateDesign_global-local}(a), is discretised in cells of $5 \times 5 \times 10$~nm$^3$, where we only explicitly include the magnetic part of the design in the micromagnetic simulations. The damping constant and the exchange stiffness are set at $\alpha =$~0.02 and $A =$~1.3~$\times$~10$^{-11}$~J/m, respectively.

\begin{figure*}
\centering{
    \includegraphics[width=0.7\textwidth]{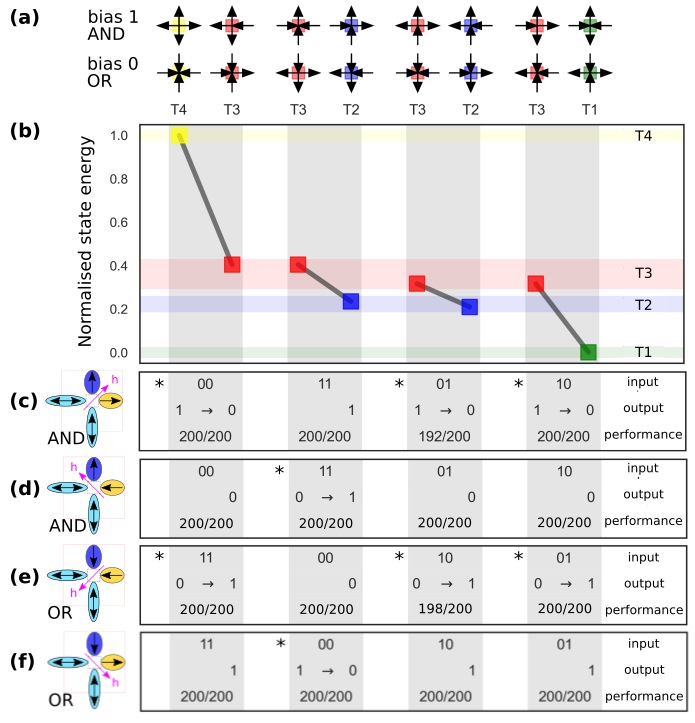}}
    \caption{(a) 16 possible spin configurations, with the bias island set to $1$ ($\uparrow$, top row) or $0$ ($\downarrow$, bottom row). (b) The normalised magnetostatic state energies (at $T=300$~K) result in an energy hierarchy $E(T4)>E(T3)>E(T2)>E(T1)$. (c-d) The logic table for the AND gate for different input configurations, with the bias island set to $1$ by the field $h$ and the output island to $1$ and $0$, respectively. The correct output is either obtained from a switching event (e.g., 1 → 0) or by the system remaining in a low-energy state. The last row gives the performance of the simulated gate, i.e. the success rate at which the correct operation (switch or stay) is observed. The operation (switch versus stay) is reversed for both AND gates, as indicated by the asterisks. (e-f) Equivalent logic table for the OR gate, with the bias island set to $0$ by the field $h$ and the output island to $0$ and $1$, respectively. These gates are related to the AND gates in (c) and (d), respectively, by a global reversal symmetry ($0$ $\leftrightarrow$ $1$).}
    \label{fig_sts}
\end{figure*}

To investigate the performance of our gate, we check whether the magnetisation of the output island is oriented along the desired direction after applying a horizontally polarised light pulse. Each input configuration is tested 200 times, with the magnetisation direction of the vertical bias island initialised upwards and downwards to recover logical AND and OR behaviour, respectively, to demonstrate the reconfigurability of our gate. In addition, to reveal underlying symmetries, we initialise the output either at bit $1$ (by applying the small external field $h$ in the second step of the field protocol to the top/bottom right) or at bit $0$ (by applying $h$ to the top/bottom left). Figures~\ref{fig_sts}(c-f) show all of the four possible gate initialisations with respect to the state of the bias and output island. In our micromagnetic simulations, these initial states are set directly instead of via a field protocol for the sake of simplicity. 

\subsection{Results}

The performance of the simulated gates, i.e., the success rate of the logically correct operations, is determined by the energy difference between the involved states. In the case of a pronounced energy difference (first, second, and last column of Fig.~\ref{fig_sts}), the performance is always 200/200. However, for a smaller energy difference (third column), the performance depends on whether a reversal of the output island is required in order to obtain the desired logical behaviour, as indicated with the asterisks.

Although the energy difference is the same for AND 01* and AND 01 [see the third column of Figs.~\ref{fig_sts}(c) and (d)], the performance is slightly worse when a switching event has to occur (192/200 versus 200/200). The symmetry breaking is explained by the {\it ringing effect} which takes place after a reversal. Due to this rotational inertia effect, the final state oscillates about the minimum energy for some time. As a consequence, the energy difference with the high-energy state, which is already very small for this input configuration, momentarily reduces, making it more likely to switch back to the logically incorrect high-energy state.

In Figure~\ref{fig_sts}, the AND gates shown in panel (c) and (d) are related to the OR gates of panel (e) and (f), respectively, by a global reversal symmetry, which allows to convert an AND gate into an OR gate and vice versa~\cite{GYP-18}. When reversing all of the magnetic moments (bits), the magnetostatic energy remains unchanged, as it scales as ${\bf m}^2$. Therefore, the performance of such globally symmetric gates is expected to be identical. As verified with micromagnetic simulations, the observed success is the same, except for AND 01* (192/200) and OR 10* (198/200), with the difference of events ascribable to Poisson statistics\footnote{Using Poisson statistics of small numbers with an average error rate of $\lambda=$5 per 200 simulations, we can calculate the probability $P(X=k) = \frac{\lambda^k}{k!} \exp{(-\lambda)}$ of having $k$ unsuccessful switching events when simulating 200 times. For AND 01* and OR 10*, this probability is equal $P(X=8)=6.5\%$ and $P(X=2)=8.4\%$, respectively. These numbers are plausible compared to $P(X=\lambda)=17.5\%$}.

\subsection{Discussion}
Due to the symmetries in the system, we can restrict the discussion to the AND gate shown in panel (c) of Fig.~\ref{fig_sts} without a loss of generality.

\begin{figure}
\centering{
    \includegraphics[width=0.5\textwidth]{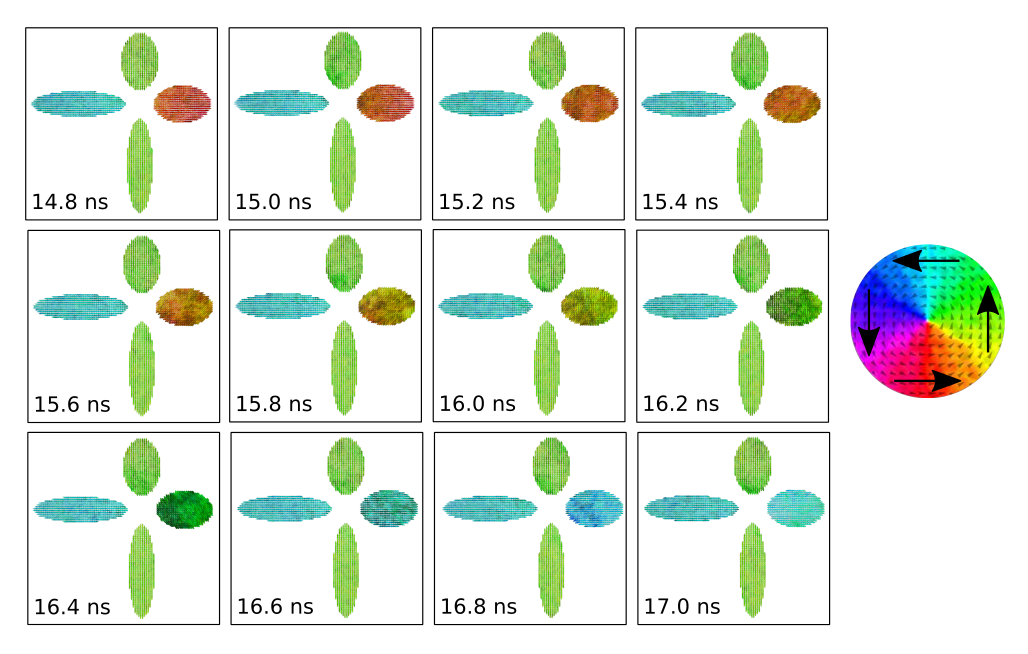}}
    \caption{Time-dependent snapshots of the AND gate with input configuration $01$ [see third column of Fig.~\ref{fig_fieldprotocol}(b)] obtained from micromagnetic simulations. The continuous change of the magnetisation orientation of the output island, represented by the colours of the colour wheel, from red (at 14.8~ns) via green (at 16.2~ns) to cyan (at 17.0~ns), indicate a coherent (counterclockwise) rotation of the net moment.}
    \label{fig_coherentREV}
\end{figure}

The performance of our gate is determined by the time it takes for the output island to reverse its magnetisation. The temperature profile shown in Fig.~\ref{fig_tempProfile}(b), where the temperature of the output island remains sufficiently high for only 10~ns, dictates that the magnetic switching needs to take place at the same timescale if the logical behaviour requires such a transition. To determine the switching time and to understand the relevant underlying physics, we use a simplified model in which transitions occur via a coherent reversal. This model is validated by the full micromagnetic simulations: snapshots of the magnetisation taken at time intervals of 0.2~ns corroborate the picture of a coherent reversal, as shown in Fig.~\ref{fig_coherentREV}.

For a transition in which the states are separated by an energy barrier $E_{{\rm bar}}$, the switching time can be expressed by an Arrhenius law as
\begin{equation}
    \tau_{s} =  \tau_0\ {\rm exp} \bigg( \frac{E_{{\rm bar}}}{k_{\rm B} T} \bigg),
    \label{eq_arrhenius}
\end{equation}

with $k_{\rm B}$ the Boltzmann constant and $\tau_0 = \frac{1}{2f_0}$ the inverse of twice the attempt frequency. In the case of nanomagnetic islands with uniaxial anisotropy, $\tau_0$ equals~\cite{BRO-59} 
\begin{equation}
    \tau_0 =  \frac{1+\alpha^2}{2\alpha \gamma } \sqrt{\frac{\pi M_{{\rm s}}^2 k_{\rm B} T}{4 K_{\rm u}^3 V}}.
    \label{eq_tau0}
\end{equation}

In this equation, $V$ is the ferromagnetic volume of the island, and $\gamma$ stands for the gyromagnetic ratio. The uniaxial anisotropy constant is determined as $K_{\rm u} = \frac{E_0}{V}$, where $E_0$ denotes the energy barrier of an isolated island, which can be estimated as the energy difference between a state with magnetisation along the geometrically short (magnetic hard) axis and a state with magnetisation along the geometrically long (magnetic easy) axis. Due to the temperature dependence of $M_{\rm s}$ and $E_0$, the value of $\tau_0$ varies upon illumination, ranging from approximately $10^{-10}$~s (lowest temperatures) to $10^{-8}$~s (highest temperatures). However, because of the exponential dependence of $\tau_{s}$ on the energy barrier, the temperature dependence of $\tau_0$ has an almost negligible effect on the total switching rate.

\begin{figure}
\centering{
    \includegraphics[width=0.45\textwidth]{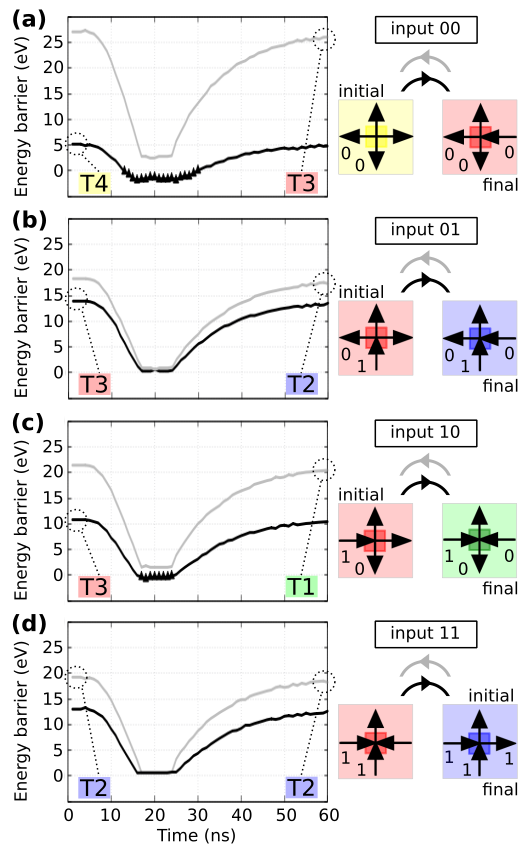}}
    \caption{An overview of the transitions that could take place during the operation of our AND gate, with input configuration (a) $00$, (b) $01$, (c) $10$, and (d) $11$. The mean-field energy barriers are determined via Eq.(\ref{eq_mfBar}) and can therefore become negative, as is the case in panel (a) and (c). The black triangles indicate when the switching time $\tau_s$ from Eq.(\ref{eq_arrhenius}) is below 1~ns, for which switching is likely to occur. For input $01$ [panel (b)], the switching times well exceed the 1~ns limit, although our simulations have shown that these transitions do take place.} 
    \label{fig_AND_trans}
\end{figure}

The energy barrier that must be overcome in the transition from state $i$ to state $j$ can be calculated with the mean-field barrier method~\cite{KOR-20} as
\begin{equation}
    E_{{\rm bar}}^{i\to j} = E_0 + \frac{1}{2}(E_{{\rm ms}}^j - E_{{\rm ms}}^i),
    \label{eq_mfBar}
\end{equation}
with $E_{{\rm ms}}$ the (time- and temperature-dependent) magnetostatic energy. In Figure~\ref{fig_AND_trans}, the energy barriers related to the transitions of our AND gate are shown as a function of time. Note that the mean-field energy barrier can be negative. Such transitions do not involve a potential well into which a state can be trapped, but should be interpreted as barrierless, where a larger negative value corresponds to a more likely transition.  

Equations~(\ref{eq_arrhenius})-(\ref{eq_tau0}) allow us to determine when the switching time is less than 1~ns. On this time scale, we expect that switching is likely to occur, since the temperature of the output island is close to $T_C$ for a period of 10~ns. However, in some cases, e.g., the AND gate with input configuration $01$ [see Fig.~\ref{fig_AND_trans}(b)], the switching times well exceed the 1~ns limit in the entire time window, although our simulations have shown that these transitions do take place.

The discrepancy between the switching time obtained by the mean-field barrier, which makes a reversal unlikely, and the micromagnetic simulations of switching events can be explained by the fact that the energy barrier used in the mean-field barrier method is independent of the magnetostatic energy of transient states, i.e., whether the magnetisation rotates clockwise or counterclockwise. It has been shown that the energy barrier for clockwise and counterclockwise rotation can differ due to the magnetic environment, leading to enhanced transition rates~\cite{LEO-21}.

\begin{figure*}
\centering{
    \includegraphics[width=\textwidth]{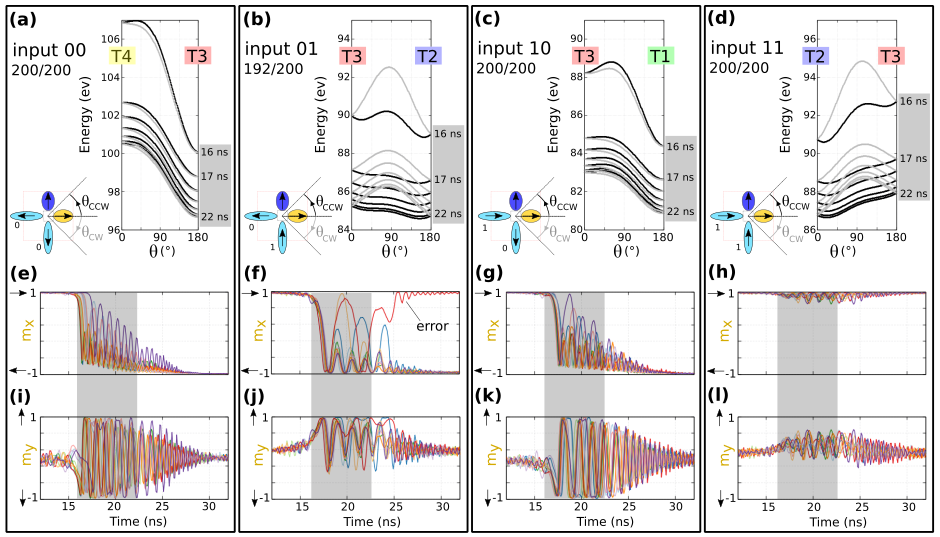}}
    \caption{(a-d) Static energy profiles $E(\theta)$ of our AND gate for a clockwise (CW, gray) and counterclockwise (CCW, black) reversal of the output island, displayed at times between 16~ns and 22~ns for which the output temperature is maximal. The energy profiles estimate the variation of the switching barriers during the thermoplasmonic heating cycle. In the case of input $01$, shown in (b), the energy profiles become almost monotonic for CCW rotations, and thus allow potential back-switching events. (e-h) and (i-l) show the simulated time evolution of the horizontal and vertical component of the magnetic moment of the output island, respectively. The results are shown for ten equivalent micromagnetic simulations run with different realisations of the thermal fluctuations.}
    \label{fig_Eprof_all}
\end{figure*}

Therefore, as shown in Fig.~\ref{fig_Eprof_all}(a-d), we calculate the magnetostatic energy of all of the transient states, based on the assumption that the nanomagnets feature quasi-uniform magnetic order, as supported by Fig.~\ref{fig_coherentREV}. In this calculation, the input and bias islands are assumed to be uniformly magnetised along their long axis, while the uniform magnetisation of the output island is set at an angle $\theta$ with respect to its easy axis, with $\theta$ ranging from $-180^{\circ}$ to $+180^{\circ}$ in steps of $1^{\circ}$. Due to the time dependence of the saturation magnetisation [see Fig.~\ref{fig_tempProfile}(c)], the energy profiles $E(\theta)$ change as a function of time. The profiles relevant to understand the observed switching events correspond to a time of about 20~ns, since the temperature of the output island then reaches a maximum, as shown in Fig.~\ref{fig_tempProfile}(b). Based on the energy profiles of Fig.~\ref{fig_Eprof_all}(a-d), we are able to explain the performance of our AND gate, in particular why input configuration $01$ can give rise to an erroneous output. 

For input configuration $01$ of our AND [Fig.~\ref{fig_Eprof_all}], the energy profiles become monotonic functions, instead of having two minima which are separated by a well-defined barrier. Therefore, it is no longer meaningful to use the Arrhenius law of Eq.~(\ref{eq_arrhenius}) to determine the switching time $\tau_s$. In the case of vanishing energy barriers, the switching time matches the time scales on which the magnetisation dynamics takes place, i.e. the precession frequency. As this zero-barrier regime is maintained for several nanoseconds, the desired transitions are likely to occur. However, during that time scale of vanishing switching barriers, the moment can return to its original orientation due to a subsequent switching event (ringing, as mentioned above). The probability of these ``erasures'' depend the energy difference $\Delta E$ between the high-energy state and the low-energy state. For input configuration $01$ [Fig.~\ref{fig_Eprof_all}(b)], the energy difference is the lowest, as a result of which the likelihood to end up in the logically incorrect state is the highest. This is illustrated in Figs.~\ref{fig_Eprof_all}(e-h), where the horizontal component of the magnetic moment of the output island, $m_x$, is plotted as a function of time. This magnetic moment, initially pointing to the right ($m_x=1$), does not change direction for input configuration $11$ [Fig.~\ref{fig_Eprof_all}(h)] and reverses reliably to $m_x = 0$ for input configuration $00$ [Fig.~\ref{fig_Eprof_all}(e)] and $10$ [Fig.~\ref{fig_Eprof_all}(g)]. This is in contrast to input configuration $01$ for which there is a nonzero probability to yield an erroneous output $m_x=1$ [Fig.~\ref{fig_Eprof_all}(f), red line]. Note that the frequency of the oscillatory behaviour can be considered as the attempt frequency, being larger for input configuration $00$ ($\approx 2$~ns) than for input configuration $10$ and $01$ ($\approx 1$~ns) due to the larger slope of $E(\theta, T)$.
    
Regarding possible reversal pathways for the output island, for input configurations $00$ and $10$, there is barely any difference between the energy profiles related to clockwise (gray) and counterclockwise (black) rotation [Fig.~\ref{fig_Eprof_all}(a) and (c)], which means that both transition paths coexist, increasing the total switching probability by a factor 2, and consequently enhancing the performance of our gate. In contrast, a pronounced difference between clockwise and counterclockwise rotation is observed for input configuration $01$ [Fig.~\ref{fig_Eprof_all}(b)]. This is because the vertical islands both have a magnetisation pointing up, giving rise to a net magnetic field with which the magnetisation of the output island tends to align. Thus, the clockwise rotation in which the magnetisation of output island opposes this field results in a higher energy barrier as compared to the counterclockwise rotation. As a result, the counterclockwise rotation is the predominant switching mechanism for input configurations $01$ [Fig.~\ref{fig_Eprof_all}(b)] and $11$ [Fig.~\ref{fig_Eprof_all}(d)], with the latter not requiring a switching event for the logically correct output. This claim is supported by Figs.~\ref{fig_Eprof_all}(i-l), which show how the vertical component of the magnetic moment of the output island, $m_y$, evolves in time. For input configuration $01$ [Fig.~\ref{fig_Eprof_all}(j)], $m_y$ is skewed towards positive values. In contrast, for input configurations $00$ and $10$ [Figs.~\ref{fig_Eprof_all}(i,k)], it oscillates between $m_y=-1$ and $m_y=1$.

Our micromagnetic simulations thus show that the gate performance can reach 100\% when its energy is significantly reduced due to the desired switching events. This is always the case, except for input configuration $01$, for which the AND gate has a 96\% performance (192/200). An improvement in the performance is achievable by increasing the energy difference between the logically correct and incorrect states. In this regard, a promising path to explore is the deformation of the vertex from square to rectangular. An alternative is to use a bias island made of permalloy instead of nickel, albeit at the cost of abolishing the reconfigurability via the light polarisation.

\section{Conclusion}
In this work, we discuss how the vertex of a square artificial spin ice, consisting of hybrid nickel-gold and permalloy-gold islands, can function as a NML gate by taking advantage of light-induced polarisation-dependent (and thus island-specific) heating. The operating speed of our thermoplasmonic NML gate exceeds that of a globally-heated thermally-driven NML gates, since the switching events take place at nanosecond time scales. The optical energy costs of operation, in the order of 100~pJ for the specific laser considered here, is furthermore on par with the energy costs of storing bits in a commercial hard disk drive (HDD) or solid-state drive (SDD), which is estimated to be in the order of 1~nJ/bit. Thus, our gate has a technologically attractive operating speed, just like clocked NML gates, and combines this advantage with the high energy-efficiency of thermally-driven gates.

Furthermore, our thermoplasmonic NML gate is reconfigurable between AND logic and OR logic operation, either by reversing the initial magnetisation direction of the bias island with an applied field, or optically, by changing the order in which horizontally and vertically polarised light illuminates the gate. 

Although the potential of optical reconfigurability of the gate operation has not been fully exploited in this work -- there are only a limited number of transitions which can take place in our simple single-vertex gate -- it can be a powerful tool in more complex gate designs. Further research will therefore focus on in-memory computing in gates which are reconfigurable in a way that cannot be achieved with fields, hence offering the ability to address the logic functionality of individual gates in systems where multiple gates are integrated.

\begin{acknowledgments}
This work was supported by the Fonds Wetenschappelijk Onderzoek (FWO-Vlaanderen) with (senior) postdoctoral research fellowships (J.L.). 
N.L., M.M., and P.V. acknowledge support from the Spanish Ministry of Science, Innovation and Universities under the Maria de Maeztu Units of Excellence Programme (MDM-2016-0618) and the pre-doctoral Grant PRE2019-088070 (M.M.), as well as from the Spanish Ministry of Science, Innovation and Universities and the European Union under the project RTI2018-094881-B-I00 (MICINN/FEDER). The work of N.L. was supported via the European Union’s Horizon 2020 research and innovation programme under the Marie Słodowska Curie Grant Agreement No. 844304 (LICONAMCO).
\end{acknowledgments}

\section*{Appendix A: Absorption cross section}
\label{appB}
\begin{figure}
\centering{
    \includegraphics[width=0.5\textwidth]{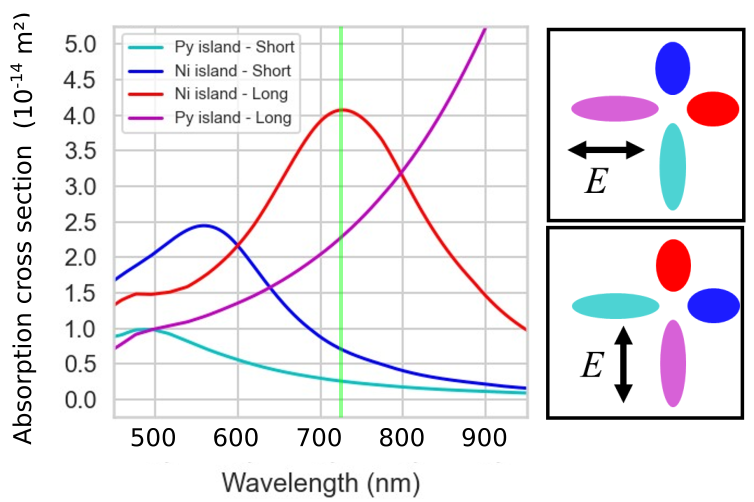}}
    \caption{Calculated absorption cross section spectra for the Au/Ni (red and blue) and Au/Py (magenta and cyan) nanoislands with light polarised as indicated by the arrows. The maximum absorption in the Ni nanoislands (output and bias) is obtained when the wavelength of the light is $\lambda=728$~nm, as indicated by the green vertical line.}
    \label{fig_absCross}
\end{figure}

The calculated absorption cross section for the nanoislands with the light polarised horizontally is shown in Fig.~\ref{fig_absCross}. To maximise the absorption displayed by the output island when the light is polarised along the island’s long axis, the wavelength of the light should be $\lambda=728$~nm. For experimental implementation of the gate, the position of the LSP absorption peak can be tuned by changing the material of the islands plasmonic layer and the thickness. Parasitic heating of the input island can be reduced by replacing Au for a non-noble-metal material, which would result in a dramatically decreased optical absorption cross section~\cite{2017Baffou}.

The spectral dependence of the polarisation-dependent absorption cross sections of the elliptical nanoislands are calculated using the method described in Ref.~\cite{MAC-13}. The geometrical dimensions are the ones proposed in the main text, (150~nm)$\times$(100~nm) for the bias/output island and (250~nm)$\times$(70~nm) for the inputs. The height is set to 20~nm and 30~nm for Ni/Py and Au, respectively. The islands are embedded in a medium with refractive index $n=1.2$, an average for the glass and air~\cite{MAC-13}. The wavelength-dependent refractive indices of gold, nickel, and permalloy are obtained from~\cite{1972johnson, 1993vivsvnovsky, 2017tikuivsis} via~\cite{rii}.

\section*{Appendix B: Saturation magnetisation and temperature}
\label{appC}
\begin{figure}
\centering{
    \includegraphics[width=0.4\textwidth]{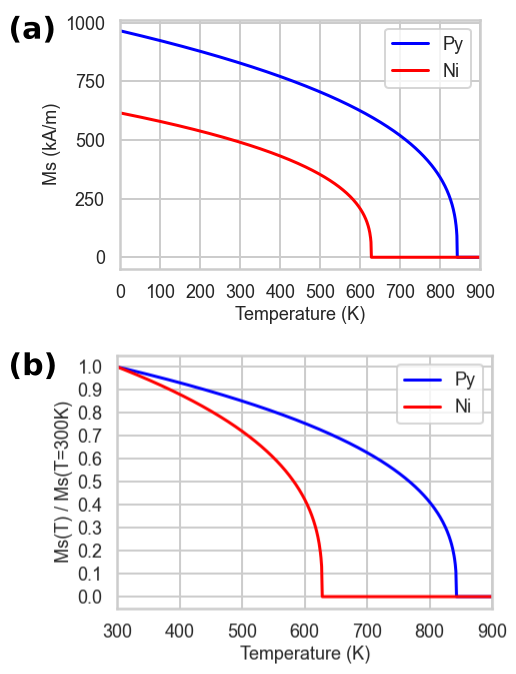}}
    \caption{(a) The saturation magnetisation as a function of temperature for Ni (red) and Py (blue). Panel (b) shows the temperature dependence of the saturation magnetisation with respect to the room temperature value.}
    \label{fig_msat-temp}
\end{figure}

The saturation magnetisation $M_{{\rm s}}(T)$ as a function of temperature, calculated as described in the main text, is shown in Fig.~\ref{fig_msat-temp}(a) for Ni and Py. Nickel has a lower saturation magnetisation and Curie temperature $T_C$ compared to permalloy. Fig.~\ref{fig_msat-temp}(b) compares the saturation magnetisation $M_{{\rm s}}(T)$ of each material to their respective value at room temperature $M_{{\rm s}}(T_{\rm RT})$. It shows the more rapid relative reduction of the relative saturation magnetisation in nickel, which at 600~K is only 42\% of the room-temperature value, while permalloy retains 75\% of $M_{\rm s}(T_{\rm RT})$. In order to achieve a reduction to 40\% of the saturation magnetisation in permalloy, the temperature would need to be 200~K higher (i.e.\ about 800~K).

\section*{Declaration of Competing Interests}
The authors declare no competing interests.

\end{document}